# Polar State in Freestanding Strontium Titanate Nanoparticles

Trevor A. Tyson[1], Tian Yu[1], Mark Croft[2], Megan E. Scofield[3], Dara Bobb-Semple[3], Jing Tao[4]
Cherno Jaye[5], Daniel Fischer[5], and Stanislaus S. Wong[3,4]

[1]Department of Physics, New Jersey Institute of Technology, Newark, NJ 07102
[2]Department of Physics and Astronomy, Rutgers University, Piscataway, NJ 08854
[3]Department of Chemistry, State University of New York at Stony Brook, Stony Brook, NY 11794
[4]Condensed Matter Physics and Materials Science Department, Brookhaven National Laboratory, Upton, NY 11973
[5]Materials Science and Engineering Laboratory, National Institute of Standards and Technology, Gaithersburg, MD 20899

Corresponding Authors:
T. A Tyson, e-mail: tyson@njit.edu
S. S. Wong, e-mail: sswong@bnl.gov

### Abstract

Monodispersed strontium titanate nanoparticles were prepared and studied in detail. It is found that ~10 nm as-prepared stoichiometric nanoparticles are in a polar structural state (with possibly ferroelectric properties) over a broad temperature range. A tetragonal structure, with possible reduction of the electronic hybridization is found as the particle size is reduced. In the 10 nm particles, no change in the local Ti-off centering is seen between 20 and 300 K. The results indicate that nanoscale motifs of $SrTiO_3$ may be utilized in data storage as assembled nano-particle arrays in applications where chemical stability, temperature stability and low toxicity are critical issues.

$SrTiO_3$ is known to be an incipient ferroelectric material in which the onset of ferroelectricity is suppressed at low temperature by quantum fluctuations and the existing soft modes of the system never become unstable [1]. $SrTiO_3$ possesses a transition to a non-ferroelectric state near 105 K, corresponding to the rotation of the $TiO_6$ polyhedra, which preserves the inversion center. The close proximity of the high purity $SrTiO_3$ (with tolerance factor t=1) to the ferroelectric state was emphasized by its conversion to this state by $^{16}O$ to $^{18}O$ isotopic substitution [2]. Under this substitution, the ferroelectric transition temperature was found to be ~23 K. In addition to isotopic substitution, substrate induced tensile strain in epitaxial films [3] can convert this system to the ferroelectric phase with a ferroelectric transition above room temperature. Chemical doping or even weak Sr/Ti off stoichiometry are also known to stabilize the ferroelectric phase [3, 4]. The possibility of integrating $SrTiO_3$, $BaTiO_3$, $PbTiO_{3,}$ as well as other simple oxides with silicon may increase the probability of merging together both microelectronics and enhanced storage capabilities [5]. The nature of ferroelectricity in $SrTiO_3$ films has been extensively studied by Raman and optical spectroscopy [6] as well structural measurements via x-ray diffraction and x-ray absorption methods [7,8, 9, 10, 11]. Films of $PbTiO_3$ and $BaTiO_3$ down to ~1 to 2 nm have been found to exhibit ferroelectricity [12]. One can envision patterned nanoscale arrays of $ATiO_3$ oxides or alternatively self-assembled nanoscale units of these oxides as information bits (1 or 0). A detailed understanding of the $SrTiO_3$ system as thin films and more importantly as nanoscale materials is required in order to enable support technological push towards small electronic devices with high storage densities.

Nanoscale properties of the $BaTiO_3$ system (and to a significantly lesser extent $SrTiO_3$) have been probed in detail and have been reviewed in recent articles (See for example Refs. [13,14]). Pair distribution function measurements on $BaTiO_3$ suggest an exponential weakening of ferroelectricity in $BaTiO_3$ with a reduction in size but no abrupt crossover [15]. The particle size and temperature dependence of the ferroelectric phase transition in nanoscale $BaTiO_3$ was studied in detail by examining local surface plasmon resonance shifts [16]. However, the exploration of possible polar structural states which would support a polar state with possible ferroelectricity in nanoparticles of the $SrTiO_3$ system has not been symmetrically performed. $SrTiO_3$ nanoparticles have been produced by wet chemical methods and physical methods such as pulsed laser vapor deposition [17]. Samples produced by ball-milling have been found to exhibit a core shell structure and broad, inhomogeneous particle size distributions, the latter of which is consistent with this specific synthesis approach [18]. In film and a few nanoparticle studies of $SrTiO_3$, structural and Raman probes reveal non-cubic structural features manifested as tetragonal lattice structure and polar Raman phonons (such as the $(TO)_2$, and $(TO)_4$ transvers optical and $(LO)_3$, $(LO)_4$ longitudinal phonons). One possible model, consistent with the $BaTiO_3$ system also, is the possibility of a polar surface (ferroelectric) and a non-polar interior to the nanoparticles.

In this work, we have synthesized single phase monodispersed nanoparticles of $SrTiO_3$ measuring approximately 10 nm, 82 nm, and 112 nm, respectively, in diameter by soft chemistry methods and compared these samples with bulk analogues. Particle size and morphology were assessed by transmission electron microscopy (TEM) measurements. We have studied them using x-ray absorption measurements which probe the behavior of entire particles including not only their interior but also their surface. Examination of the pre-edge spectra reveals that the ~10 nm $SrTiO_3$ sample possesses significant off-centering of Ti. Extraction of the radial distribution of the Ti-O bonds shows a large bond asymmetry, consistent with a polar structural state. Raman measurements of this sample reveal the existence of prominent polar phonons. The results indicate that the ~10 nm particles are in a polar state over the temperature ranges studied (20 to 300 K). No variation of the near edge peak intensity with temperature has been found, further supporting the stability of the polar phase in these ~10 nm nanoparticles.

Micron scale bulk samples of $SrTiO_3$ were obtained from Alfa Aesar. Nanoscale powders measuring 10.1 ± 1.1 nm, 82.6 ± 9.1 nm, and 112.5 ± 14 nm in diameter (see Fig. 1) were prepared mainly according to the methodology highlighted in Refs. [19]. Specifically, $SrTiO_3$ nanoparticles measuring approximately 10 nm in diameter were prepared utilizing a hydrothermal technique. In particular, titanium bis(ammonium lactate)dihydroxide (TALH) and strontium hydroxide octahydrate solutions were mixed in a 1: 1 molar ratio with the pH of the solution adjusted to 13.5.

Hydrazine and oleic acid were added to an autoclave with the precursor solution and heated to 120°C for 24 hrs. The sample was then isolated upon centrifugation and washed with water followed by ethanol. The 80 nm and 112 nm diameter $SrTiO_3$ nanoparticles were prepared using a different method. First, strontium oxalate, anatase titanium dioxide, and sodium chloride (1: 1: 20 molar ratio) were ground with a mortar and pestle for 15 min for the ~112 nm sample and 30 min for the 80 nm sample, respectively, until the mixture was rendered homogeneous. Nonylphenyl ether (NP-9 with a molar ratio of 3) was then added to the mixture and further ground until the mixture was uniform and homogeneous. The materials were then loaded into a porcelain reaction boat and placed in a tube furnace for 3.5 hrs at 820°C. Both samples were subsequently centrifuged and then washed with water followed by ethanol to remove any excess salt.

Electron diffraction experiments were carried out on a JEOL 2100F transmission electron microscope (TEM) at the Center for Functional Nanomaterials, Brookhaven National Laboratory. For XAFS measurements, polycrystalline samples were prepared by sieving the materials (500 mesh) and brushing them onto Kapton tape. Layers of tape were stacked to produce a uniform sample for transmission measurements with jump measured at ~1. Spectra were collected at the NSLS beamline X19A (room temperature) and as a function of temperature at X3A (~10 nm sample) at Brookhaven National Laboratory. Spectra were measured over the range 4,816 to 6,200 eV (about he Ti K-Edge at 4966 eV). Three to four scans were taken at each temperature. The uncertainty in temperature is < 0.2 K. A Ti foil reference was employed for energy calibration. The reduction and analysis of the X-ray absorption fine-structure (XAFS) data was performed using standard procedures [20]. To treat the atomic distribution functions on equal footing at all temperatures, the spectra were modeled in R-space by optimizing the integral of the product of the radial distribution functions and theoretical spectra with respect to the measured spectra. Specifically, the experimental spectrum is modeled by $\chi(k) = \int \chi_{th}(k,r)\, 4\pi r^2 g(r) dr$ where $\chi_{th}$ is the theoretical spectrum and g(r) is the real space radial distribution function based on a sum of Gaussian functions (χ(k) is the measured spectrum) [21] at each temperature. The k-range 2.80< k < 13.3 $Å^{-1}$ and the R-range 0.69< R <2.35 Å were used with $S_0^2 = 0.90$ (correction for inelastic losses). In the model, coordination numbers for the atomic shells were fixed to reflect the crystallographic values. The theoretical x-ray near edge spectra (XANES) presented were computed, as in Ref. [22] using the RELXAS programs [23]. The potentials were computed for the perovskite structures, based on clusters containing the first 110 atoms surrounding the Ti site. Ti off-centering was accommodated by coherently displacing all Ti ions by 0, 0.10, 0.15, and 0.20 Å along the z axis while holding the other atoms fixed as in Ref. [10]. For the random structure model, the x, y and z coordinates of each atoms in the cluster (111 atoms) were randomly displaced with an amplitude between 0 and 0.10 Å. Fluorescence Sr L1-Edge XANES data were collected at NSLS beamline X19A. Fluorescence Ti L2,L3-Edge data and O K-Edge data were collected at NSLS beamline U7A. Scans at the C K-Edge revealed no incorporation of carbon into the samples. Raman Spectra of both the ~10 nm and bulk sample of $SrTiO_3$ were measured with an excitation wavelength of 532 nm in backscattering geometry using a Thermo Scientific DXR Raman Microscope. High resolution x-ray powder diffraction measurements with $\lambda$ = 0.7792947 Å were collected a NSLS beamline X14A using a 1.4 m sample to detector distance.

In Fig. 1, we show TEM images of the nanoparticle powders (i.e. measuring 10.1 ± 1.1 nm, 82.6 ± 9.1 nm, and 112.5 ± 14 nm, respectively) noting that the scale for the ~10 nm sample is different from that of the other samples. Detailed studies indicate that the particles are single phase, as revealed by both X-ray diffraction and electron diffraction measurements. The morphology of the particles is cube-like with a tight variation in particle size of approximately 10%.

To understand the physical properties of the particles, X-ray absorption measurements were conducted. Fig. 2(a) shows the near edge structure of bulk samples as compared with their nanoscale analogues. Note that the region above ~ 4985 eV in the main line possesses multiple, well-defined multiple peaks which are present in both the bulk and nanoscale samples. The main peak corresponds to the 1s → 4p transitions on the Ti sites and the peaks above the main peaks are continuum resonances, corresponding to transitions from the 1s → (highly localized final states) or alternatively as contributions to the cross section from multiple scattering processes of high order. These features will be absent in samples which are amorphous as was found in the case with amorphous $SrTiO_3$ [9]. By contrast, our results indicate a high level of crystalline order present in the samples and confirm both the X-ray diffraction and electron diffraction measurements taken on these materials.

To identify the nature of the local structure about the Ti site and to explore the possibility of a polar structural state, the near edge features were examined (Fig. 2(b)). The pre-edge features near ~2970 eV correspond to a transition from Ti 1s →Ti 4p hybridized with Ti 3d and O 2p (analogous to the well-studied perovskite manganites [24]). These final states are labeled as either $t_{2g}$ or $e_g$ bands. The ($d_{x^2-r^2}$, $d_{y^2-z^2}$) orbitals always align along the Ti-O bond directions (as the Ti $p_x$, $p_y$, and $p_z$ also do) while the $t_{2g}$ orbitals ($d_{xy}$, $d_{xz}$, and $d_{yz}$) are off axis. The 3d state is split into a lower $t_{2g}$ and upper $e_g$ band by the octahedral field [25] of the lattice. The first two bands are the $e_g$ and $t_{2g}$ bands seen on the pre-edge features. The $e_g$ peak in the XANES spectra of $ATiO_3$ perovskite systems is extremely sensitive to local distortions (i.e. coordination of the Ti sites) [11], with the integrated intensity proportional to the mean square displacement of the Ti atom off the ideal site. Examination of Fig. 2(b) pre-edge peaks reveals a large enhancement of this feature in the ~10 nm sample as compared with all other samples. In Fig. 3(a), we show the pre-edge features for $PbTiO_3$, $BaTiO_3$, $CaTiO_3$, and $SrTiO_3$, supporting this model.

To understand the nature of the distortion, we computed spectra for the undistorted $SrTiO_3$ case, cases with 0.10, 0.15, and 0.20 Å Ti off-centering, respectively, as well as a model case with random displacement (0.1 A amplitude) in the x, y, and z-directions on all sites so as to mimic structural disorder (See Fig. 3(b)). Ti off-centering with a large amplitude produces an asymmetric enhanced broadened peak with its maximum shifted to lower energy as compared with the undistorted system while the random disorder model gives a narrow peak with a high degree of symmetry but a large amplitude. Weaker Ti off-centering give peaks which are progressively broadened and shifted to lower energy, as in the case of strong off-centering, but with reduced asymmetry. Applying these observations to the nanoparticle data in Fig. 2(b) suggests that modeling with Ti-off centering matches our observations. This result is also consistent with the lower amplitude of the $e_g$ peak as compared with those of $PbTiO_3$ and $BaTiO_3$ in Fig. 4. We note that both a large $e_g$ amplitude and asymmetry in the pre-edge peak are necessary indicators of Ti off-centering.

Additional spectroscopic and structural measurements reveal that the nanoparticles samples were stoichiometric. See supplementary material [26]. We conducted Sr L1-Edge, soft x-ray Ti L2/L3-Edge and oxygen K-edge synchrotron based measurements. In addition high resolution x-ray diffraction measurements were conducted to probe for tetragonal distortions. Total scattering (Bragg+Diffuse) hard x-ray pair distribution function measurements (NSLS X17A) were conducted to examine the Sr:Ti ratio independently. The results indicate that nanoparticles are stoichiometric, with reduction in hybridization of the O p and Ti d bands and O p with Sr d bands as particle size is reduced and reveal that the 10 nn sample exhibits a non-negligible tetragonal splitting with a = 3.9142(3) Å and c = 3.9118(6) Å.

Examination of the XAFS high energy part of the data can provide additional information. In Fig. 4(a), we show the XAFS structure function for the bulk and nanoparticle samples and an example of a typical fit (Fig. 4(b)). The Ti-O peaks for the nanoparticles are similar but the higher order peaks (corresponding to Ti-Sr, Ti-Ti, and multiple scattering contributions such as Ti-O-Ti forward scattering) are significantly suppressed in the ~10 nm sample. We fit the first peak in the XAFS data as the sum of two Gaussians for all nanoparticle samples (and a single peak for the bulk sample) and have extracted the radial distribution functions, g(r,) which is displayed in Fig. 4(c). A high degree of asymmetry can be observed for the case of the 10 nm sample, as expected for a ferroelectric (polar sample). We note that a sample with random disorder in O ions will possess weak high order shells (see Ref [10]) but this is not the case herein. To separate out random structural disorder from ordered local distortions supporting a polar state, additional direct structural and Raman measurements were performed.

To further confirm that the ~10 nm sample was indeed polar (ferroelectric), we measured the room temperature Raman spectra of this sample and compared it with the bulk sample (Fig. 5(a)). Indeed one can clearly see new peaks corresponding to the polar modes $((TO)_2$, $(TO)_4$, $((LO)_3$, and $(LO)_4)$.

Hence, the ~10 nm nanoparticles are in a polar state with possible ferroelectric properties. To understand the thermal stability of the polar structural state, the temperature dependence of the pre-edge features was explored. In Fig.5(b), we see that there is no strong variation of the intensity of the $e_g$ feature between 20 K and 300 K with our data showing that the polar state in the ~10 nm particles is stable over this broad range of temperatures. This should be contrasted with the case of the unstrained film (fig. 5(c)) which possesses slight Ti/Sr off stoichiometry. In the case of these films, strong polar Raman lines emerge only below ~200 K. By contrast, we note the sensitivity of the pre-edge feature (Fig. 5(c)) to the local distortion about the Ti site for bulk $SrTiO_3$. As the temperature is reduced, the peak height is suppressed, showing that the polar state supporting ferroelectricity of these systems is progressively quenched by the quantum fluctuations.

Monodispersed $SrTiO_3$ nanoparticles were prepared and studied. It is found that nanoparticles approximately 10 nm in size are in a polar structural state and possibly ferroelectric over a broad temperature range. A tetragonal distortion occurs with expansion of the lattice as the particle size is reduced. Reduced electronic hybridization occurs with size reduction. The mechanism driving these changes with particle size are not yet understood, but the results may encourage more detailed experimental work and as well as theoretical investigations. No change in the local Ti-off centering is seen between 20 and 300 K. The results indicate that nanoscale motifs of $SrTiO_3$ may be utilized in data storage as assembled arrays of high density bits.

## Acknowledgments

This work is supported in part by DOE Grant DE-FG02−07ER46402 (TAT, TY). Research (including support for MES and SSW) was supported by the U.S. Department of Energy, Basic Energy Sciences, Materials Sciences and Engineering Division and was conducted at Brookhaven National Laboratory, which is supported by the U.S. Department of Energy under Contract No. DE-AC02-98CH10886. Synchrotron powder X-ray diffraction and X-ray absorption data acquisition were performed at Brookhaven National Laboratory's National Synchrotron Light Source (NSLS) which is funded by the U.S. Department of Energy.  We thank Dr. Yuqin Zhang (NJIT) for conducting the Raman measurements on the samples.

Fig. 1. Tyson *et al.*

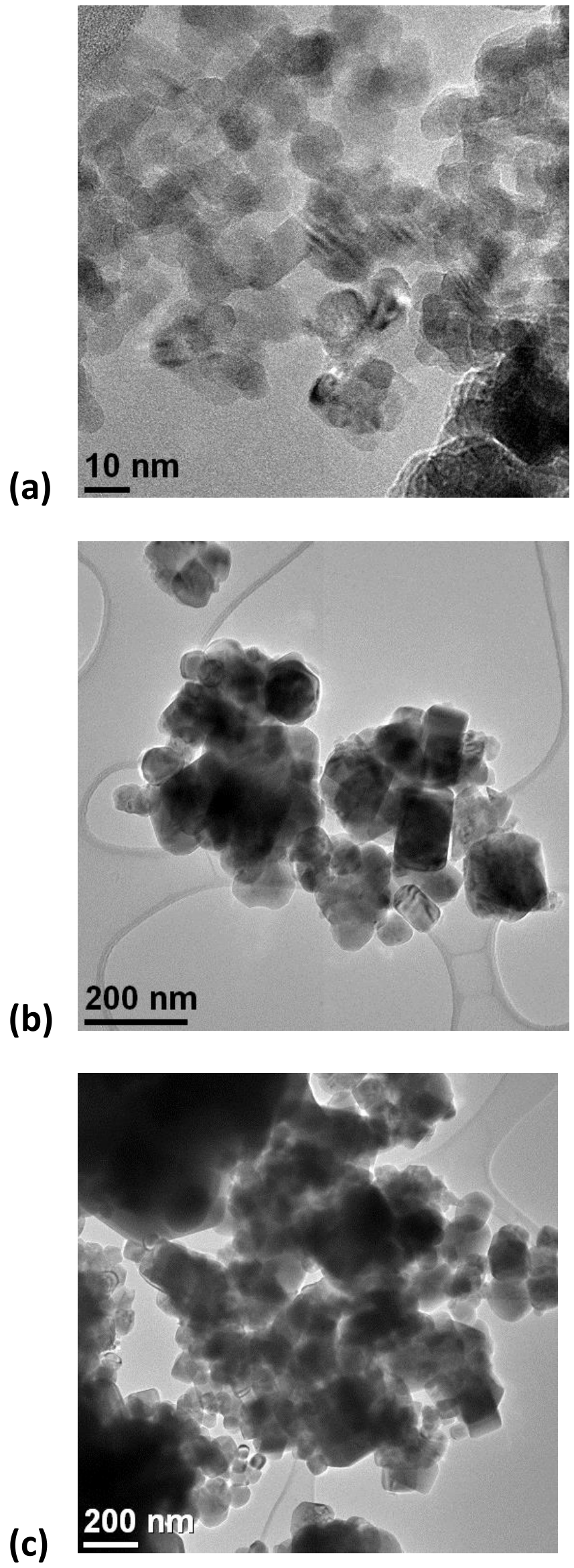

**Fig. 1**. TEM images of the (a) 10 nm, (b) 82 nm, and (c) 112 nm samples, showing the small variation in particle size for each particle set. Note that the scale bar corresponds to 200 nm for the 82 nm and 112 nm sample and 10 nm for the 10 nm sample.

Fig. 2. Tyson *et al.*

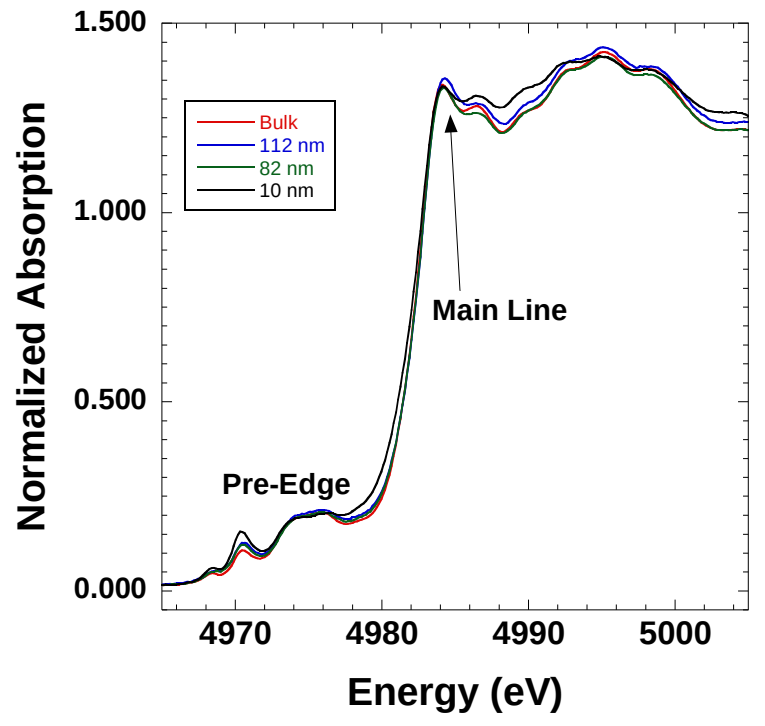

(a)

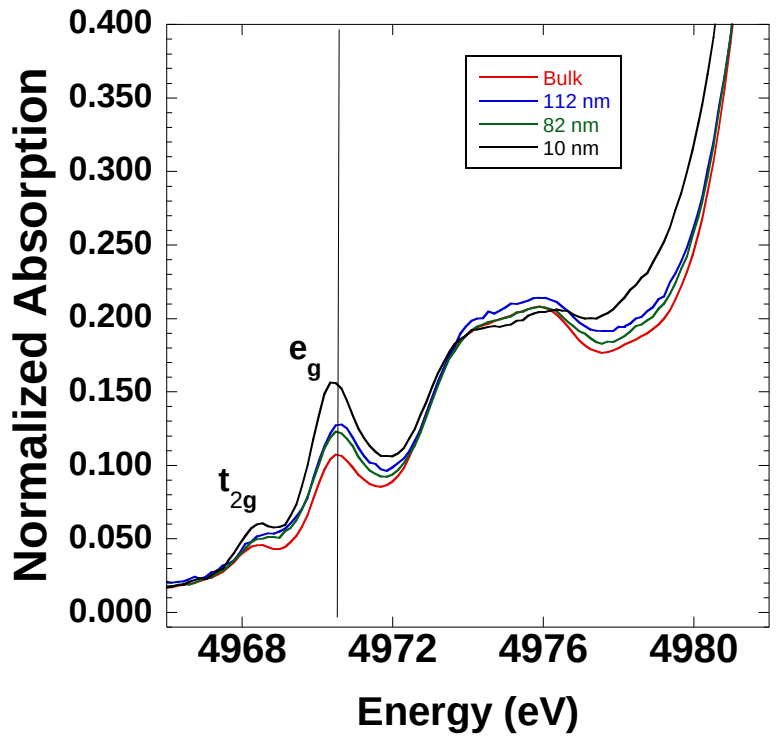

(b)

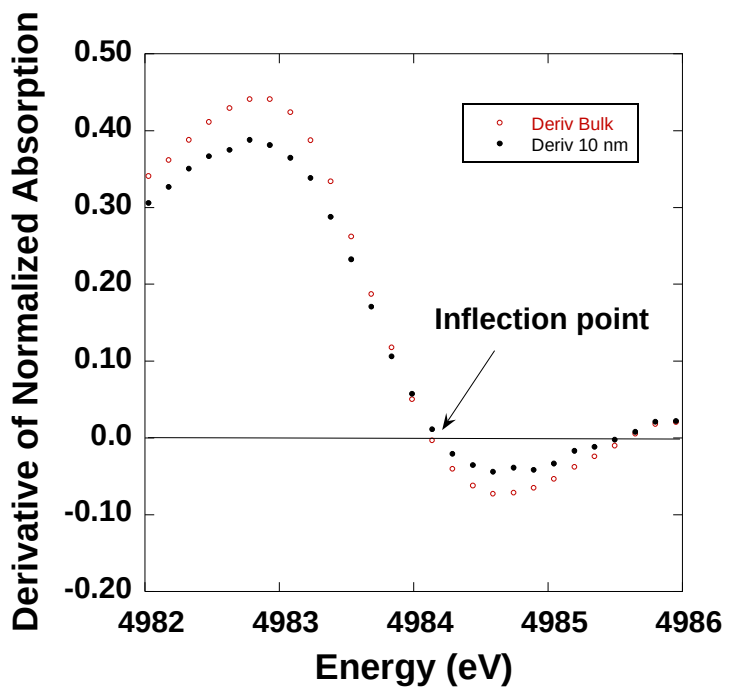

(c)

**Fig. 2**. (a) Ti K-edge XANES measurements of 10 nm, 82 nm, 112 nm, and bulk (micron scale) particles and (b) the expanded pre-edge region. Note the strong variation of the intensity of the peak labeled B with particle size. Increasing distortion enhances the peak height, and large distortion shifts it asymmetrically to lower energy. Note that even in the 10 nm sample, the features above the main line are well resolved, indicating well defined structural shells about Ti sites for all samples. (c) Inflection point of the 1 0nm and bulk sample (derivative near the main peak) in (a) showing no significant Ti ion chemical shift (see supplementary data [26]).

Fig. 3. Tyson *et al.*

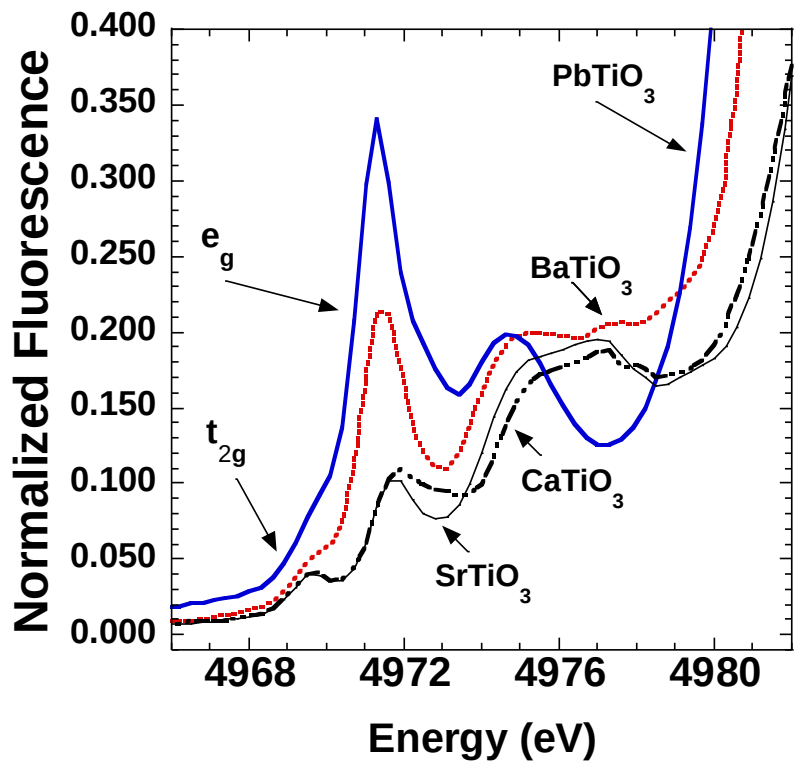

(a)

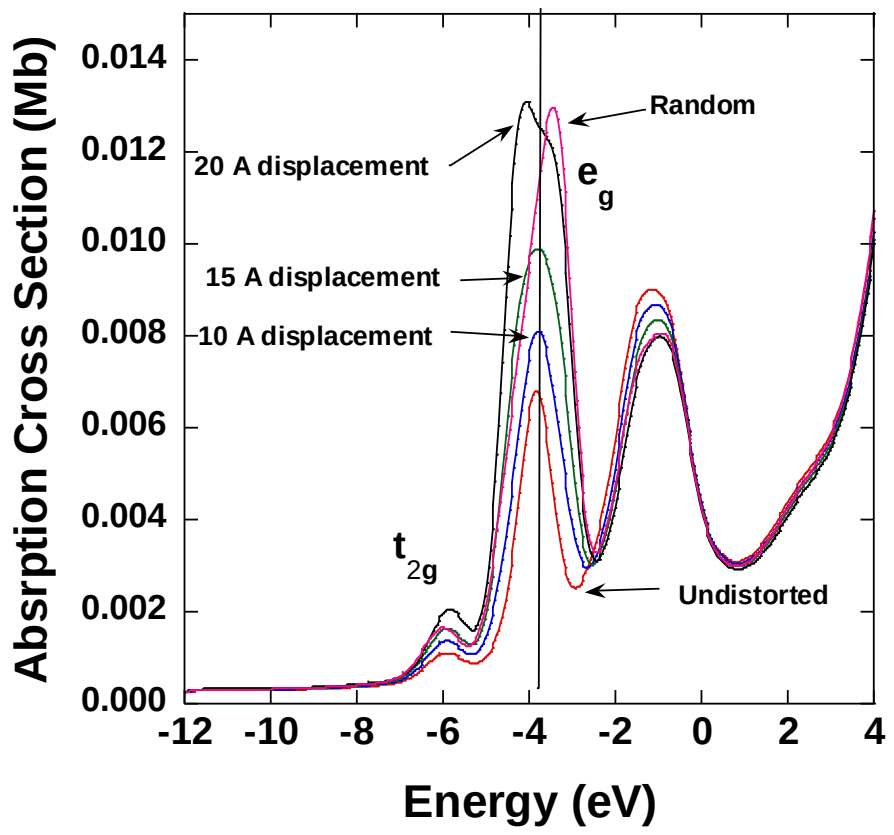

(b)

**Fig. 3**. (a) Ti K-edge XANES pre-edge regions for bulk phases of $PbTiO_3$ ($T_c$ = 763 K)), $BaTiO_3$ ($T_c$ = 493K), $CaTiO_3$, and $SrTiO_3$. (b)Ti K-edge Multiple scattering simulations of the pre-edge XANES region, modeled with random atomic displacements and with coherent 0.10 Å, 0. 15 Å, and 0.20Å z-axis atomic displacements of the Ti ions as compared with the undistorted system. Note that the random distortion model gives a symmetric peak with a position above the undistorted model, while the case for the coherent Ti off centering yields a peak shifted below the undistorted case.

Fig. 4. Tyson *et al.*

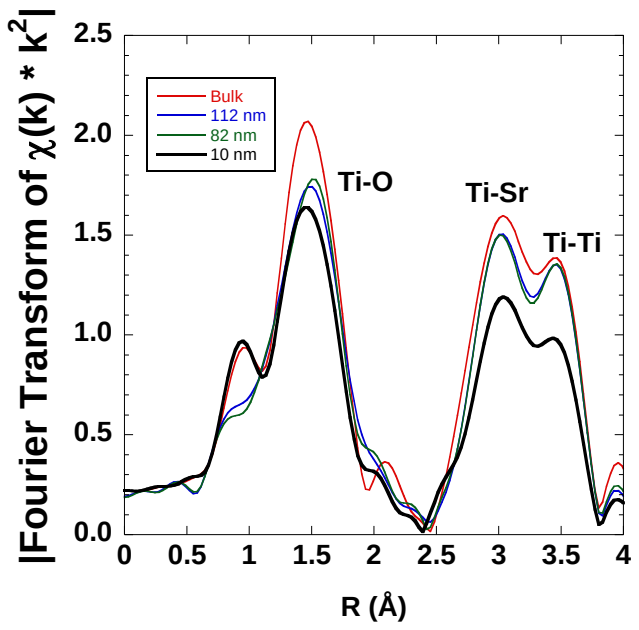

**(a)**

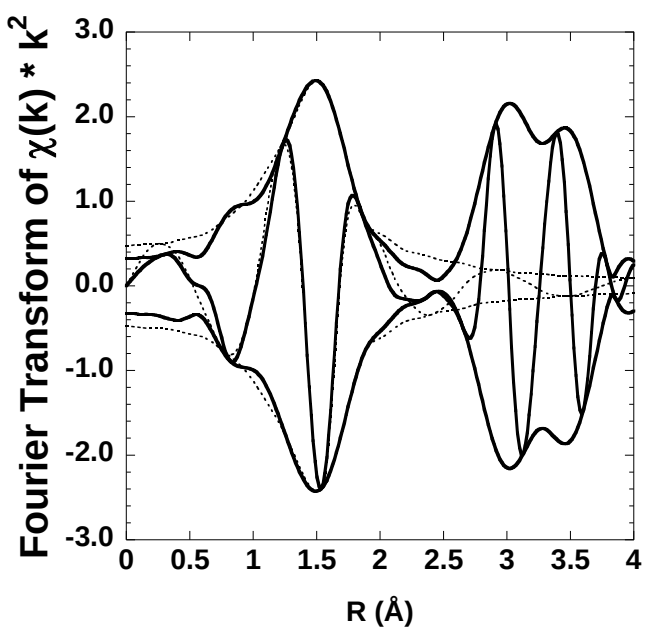

**(b)**

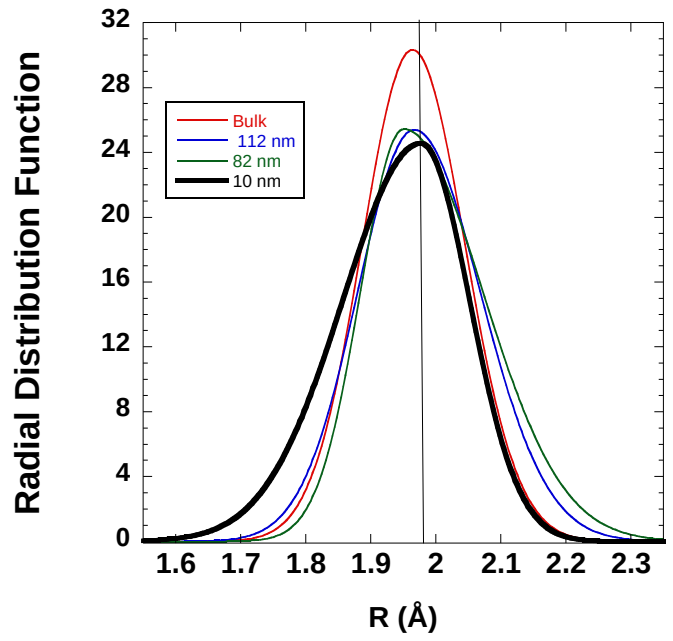

**(c)**

**Fig. 4**. (a) Fourier transform of the XAFS data for the samples. We note the significant reduction of the Ti-Sr/Ti-Ti peaks for the 10 nm sample. (b) Representative fit of the first-shell Ti-O peak for the bulk sample. (c) Radial distribution function for the first shell Ti-O peak extracted from the fitting for all four samples. We note the high degree of asymmetry in the 10 nm sample consistent with a polar structure.

Fig. 5 Tyson *et al.*

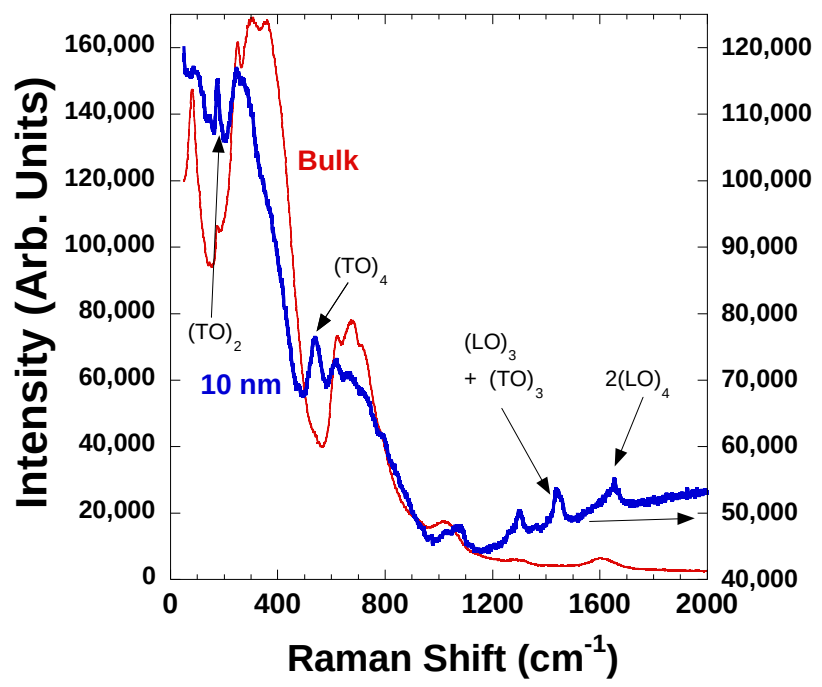

**(a)**

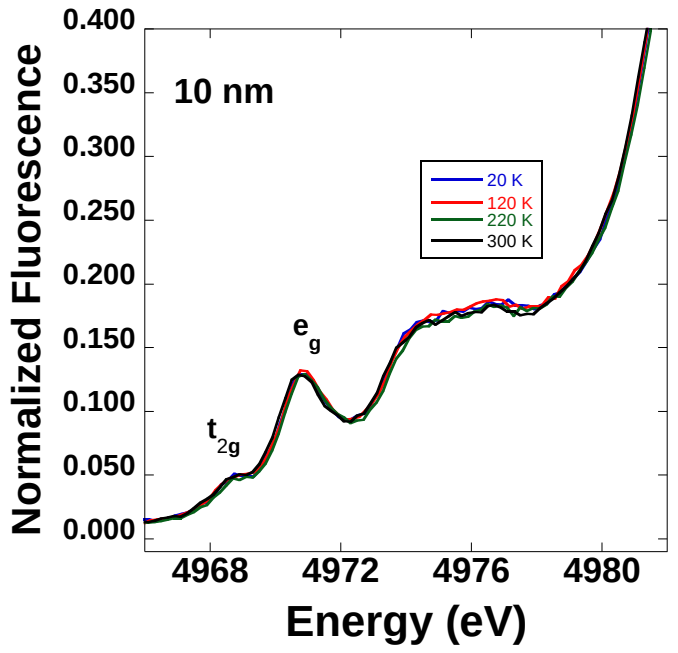

**(b)**

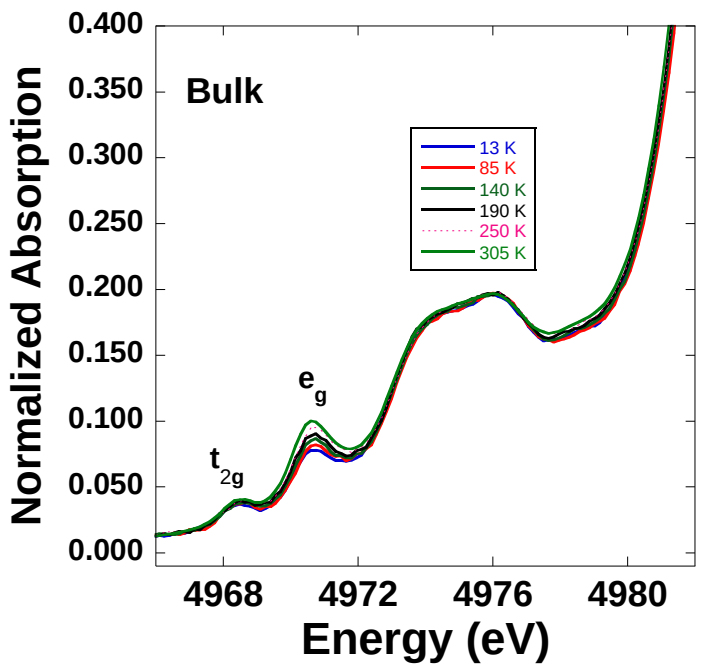

**(c)**

**Fig. 5.** (a) Raman spectra for Bulk $SrTiO_3$ and the 10 nm sample. We note the existence of polar modes $(TO)_2$, $(LO)_3$, and $(TO)_4$ in the 10 nm sample. The peaks above ~800 $cm^{-1}$ are due to two phonon excitations. Temperature dependent pre-edge features the 10 nm sample (b) and for a bulk sample (c), showing that there is no reduction in the local structural distortion with reduced temperature for the 10 nm sample, while suppression of the distortion occurs at low temperature in bulk $SrTiO_3$.

# References

[1] (a) M. E. Lines and A.M.Glass. *Principles and Applications of Ferroelectrics and Related Materials,* (Oxford Univ. Press, Oxford, 1977).

(b) K. M. Rabe, C. H. Ahn, and M. Triscone, *Physics of Ferroelectrics: A Modern Perspective* (Springer-Verlag, Berlin , 2007).

[2] M. Itoh, R. Wang, Y. Inaguma, T. Yamaguchi, Y.-J. Shan, and T. Nakamura, Phys. Rev. Lett. 82, 3540 (1999). M. Itoh, T. Yagi, Y. Uesu, W. Kleemann and R. Blinc, Sci. Technol. Adv. Mat. **5** (4), 417 (2004).

[3] T. Mitsui and W. B. Westphal, Phys. Rev. **124**, 1354 (1961).

[4] (a) Y. Y. Guo, H. M. Liu, D. P. Yu and J. M. Liu, Phys Rev B **85**, 104108 (2012).

(b) D. A. Tenne, A. K. Farrar, C. M. Brooks, T. Heeg, J. Schubert, H. W. Jang, C. W. Bark, C. M. Folkman, C. B. Eom and D. G. Schlom, Appl. Phys. Lett. **97**, 42901 (2010).

(c) H. W. Jang, A. Kumar, S. Denev, M. D. Biegalski, P. Maksymovych, C. W. Bark, C. T. Nelson, C. M. Folkman, S. H. Baek, N. Balke, C. M. Brooks, D. A. Tenne, D. G. Schlom, L. Q. Chen, X. Q. Pan, S. V. Kalinin, V. Gopalan and C. B. Eom, Phys. Rev. Lett. **104**, 197601 (2010).

[5] (a) C. Dubourdieu, J. Bruley, T. M. Arruda, A. Posadas, J. Jordan-Sweet, M. M. Frank, E. Cartier, D. J. Frank, S. V. Kalinin, A. A. Demkov and V. Narayanan, Nature Nanotechnology **8**, 748 (2013).

(b) C. S. Hellberg, K. E. Andersen, H. Li, P. J. Ryan and J. C. Woicik, Phys Rev Lett**108,** 166101 (2012).

(c) M. P. Warusawithana, C. Cen, C. R. Sleasman, J. C. Woicik, Y. L. Li, L. F. Kourkoutis, J. A. Klug, H. Li, P. Ryan, L. P. Wang, M. Bedzyk, D. A. Muller, L. Q. Chen, J. Levy and D. G. Schlom, Science **324** (5925), 367 (2009).

[6](a) H. W. Jang, A. Kumar, S. Denev, M. D. Biegalski, P. Maksymovych, C. W. Bark, C. T. Nelson, C. M. Folkman, S. H. Baek, N. Balke, C. M. Brooks, D. A. Tenne, D. G. Schlom, L. Q. Chen, X. Q. Pan, S. V. Kalinin, V. Gopalan and C. B. Eom, Phys. Rev. Lett. **104** , 197601 (2010).

(b) D. A. Tenne, X. Xi, J. Am. Ceram. Soc.**91**, 1820 (2008).

(c) S. B. Shi, W. Z. Shen and H. Wu, Appl. Phys. Lett. **91**, 112910 (2007).

(d) H. L. Yu, Y. U. Wu, X. F. Jiang, M. Q. Cai, L. P. Gu and G. W. Yang, J. Appl. Phys. **114** , 173502 (2013).

(e) S. Gupta and R. S. Katiyar, J. Raman Spectroscopy **3,** 885 (2001).

(f) A. A. Sirenko, I. A. Akimov, C. Bernhard, A. M. Clark, J. H. Hao, W. D. Si and X. X. Xi, Aip. Conf. Proc. **535**, 201-217 (2000).

(g) S. S. Majanovic and Z. V. Popovic, Solid State Phenom **61-2**, 309 (1998).

[7] (a) W. J. Maeng, Y. I. Jung and J. Y. Son, Solid State Comm. **152** , 1256 (2012).

(b) D. P. Kumah, J. W. Reiner, Y. Segal, A. M. Kolpak, Z. Zhang, D. Su, Y. Zhu, M. S. Sawicki, C. C. Broadbridge, C. H. Ahn and F. J. Walker, Appl. Phys. Lett. **97**, 251902 (2010).

(c) J. C. Woicik, H. Li, P. Zschack, E. Karapetrova, P. Ryan, C. Ashman and C. S. Hellberg, Phys. Rev. B **73**, 024112 (2006).

(d) F. S. Aguirre-Tostado, A. Herrera-Gomez, J. C. Woicik, R. Droopad, Z. Yu, D. G. Schlom, P. Zschack, E. Karapetrova, P. Pianetta and C. S. Hellberg, Phys Rev B **70**, 201403 (2004).

[8] (a) J. P. Itie, A. M. Flank, P. Lagarde, S. Ravy and A. Polian, NatoSci Peace Sec B, 51-67 (2010). A. Kodre, I. Arcon, J. P. Gomilsek and B. Zalar, X-Ray Absorption Fine Structure-XAFS13 **882**, 481 (2007).

(b) B. Ravel and E. A. Stern, Physica B **208-209,** 316 (1995)

(c) B. Rechav, Y. Yacoby, E. A. Stern, J. J. Rehr and M. Newville, Phys. Rev. Lett. **72** (9), 135 (1994).

[9] A. I. Frenkel, D. Ehre, V. Lyahovetskaya, L. Kanner, E. Wachtel and I. Lubomirsky, Phys. Rev. Lett. **99**, 215502 (2007).

[10] V. Vedrinskii, V. L. Kraizman, A. A. Novakovic, P. V. Demekhin and S. V. Urazhdin, J. Phys, Cond. Mat **10** (42), 9561 (1998).

[11] J. C. Woicik, E. L. Shirley, C. S. Hellberg, K. E. Andersen, S. Sambasivan, D. A. Fischer, B. D. Chapman, E. A. Stern, P. Ryan, D. L. Ederer and H. Li, Phys Rev B **75**, 140103 (2007).

[12] (a) D. D. Fong, G. B. Stephenson, S. K. Streiffer, J. A. Eastman, O. Auciello, P. H. Fuoss and C. Thompson, Science **304**, 1650 (2004).

(b) Tenne, P. Turner, J. D. Schmidt, M. Biegalski, Y. L. Li, L. Q. Chen, A. Soukiassian, S. Trolier-McKinstry, D. G. Schlom, X. X. Xi, D. D. Fong, P. H. Fuoss, J. A. Eastman, G. B. Stephenson, C. Thompson and S. K. Streiffer, Phys. Rev. Lett. **103**, 177601 (2009).

[13] M. Polking, M.-G. Han, A. Yourdkhani, V. Petkov, C. F. Keselowski, V. V. Volkov, Y. Zhu, G. Caruntu, A. P. Alivisatos and R. Ramesh, Nature Materials **11**, 700 (2012).

[14] (a) J. Varghese, R. W. Whatmore, J. D. Homes, J. Mater.Chem. C **1**, 2618 (2013).

(b) T. Lee and A. Aksay, Crystal Growth and Design **1**, 501 (2001).

[15] V. Petkov, V. Buscaglia, M. T. Buscaglia, Z. Zhao and Y. Ren, Phys. Rev. B 78, 054107 (2008).

[16] D. Szwarcman, D. Vestler and G. Markovich, ACSnano **5**, 507 (2011).

[17] (a) J. M. Kiat, C. Bogicevic, P. Gemeiner, A. Al-Zein, F. Karolak, N. Guiblin, F. Porcher, B. Hehlen, L. Yedra, S. Estrade, F. Peiro and R. Haumont, Phys Rev B **87** 024106 (2013).

(b) L. H. Hu, C. D. Wang, S. Lee, R. E. Winans, L. D. Marks and K. R. Poeppelmeier, Chem Mater **25**, 378 (2013).

(c) C. Yuan, S. Ye, B. Xu, W. Li, Appl. Phys. Lett. **101**, 071909 (2012).

(d) M. Makarova, A. Dejneka, J. Franc, J. Drahokoupil, L. Jastrabek and V. Trepakov, Opt Mater **32** (8), 803-806 (2010).

(e) N. Yongvanich and P. Visuttipitukkul, Adv. Mat. Res. **93-94**, 471-474 (2010).

(f) J. Xie, H. Y. Mang, X. H. Ou-Yang, T. H. Ji, Z. Y. Xiao and J. Y. Sun, J. Inorg. Mater. **23** (2), 262 (2008).

(g) Y. Okumura, C. Oi, W. Sakamoto and T. Yogo, J. Mater. Res. **23** (1), 127 (2008).

(h) F. Voigts, T. Damjanovic, G. Borchardt, C. Argirusis and W. Maus-Friedrichs, J. Nanomater. **2**, 63154 (2006).

[18] M. Sopicka-Lizweed, High Energy Ball Milling (Woodhead, Cambridge, 2010).

[19] K. Fujinami,K. Katagiri, J. Kamiya, T. Hamanaka, K. Koumoto, Nanoscale **2**, 2080 (2010).

[19] Y Mao, S. Banerjee, S.S. Wong, J. Amer. Chem. Soc. **125**, 15718 (2003).

[20] (a) B. Ravel and M. Newville, R. J. Synchrotron Rad. **12**, 537. (2005).

(b) X-Ray Absorption: Principles, Applications, Techniques of EXAFS, SEXAFS and XANES; Konningsberger, D. C., Prins, R., Eds.; Wiley: New York, 1988.

[21] K.V. Klementyev, J. Phys. D **34**, 209 (2001).

[22] Q. Qian, T. A. Tyson, C.-C.Kao, M. Croft, S.-W. Cheong, and , M. Greenblatt, Rev. B **64**, 22430 (2001).

[23] (a) T. A. Tyson, (unpublished). (b) T. A. Tyson, Phys. Rev. **49**, 12578 (1994).

[24] (a) Q. Qian, T. A. Tyson, M. Deleon, C. C. Kao, J. Bai, and A. I. Frenkel, Journal of Physics and Chemistry of Solids **68**, 458 (2007).

(b) Q. Qian, T.A. Tyson, C.-C. Kao, M. Croft, A.Y. Ignatov, Appl.Phys. Lett.**80**, 3141 (2002).

(c) Q. Qian, T.A. Tyson, S. Savrassov, C.C. Kao, M. Croft, Phys. Rev. B. **68**, 014429 (2003).

[25] (a)   . V. Von Benthem, C. Elsasser and  R. H. French, J. Appl. Phys. **90**, 6156 (2001). Von Benthem, C. Elsasser and  R. H. French, J. Appl. Phys. **90**, 6156 (2001).

(b) see also the case of $BaTiO_3$, L. A. Chasse, S. Borek, K. M. Schindle, M. Trautmann, M. Huth and S. Steudel, Phys. Rev. B **84**, 195135 (2011).

[26] See supplementary material at [URL will be inserted by AIP] for detailed synchrotron based spectroscopic and diffraction analysis of the samples showing negligible atomic defects and tetragonal distortion in the 10 nm sample.

# Supplementary Data

## Polar State in Freestanding Strontium Titanate Nanoparticles

Trevor A. Tyson[1], Tian Yu[1], Mark Croft[2], Megan E. Scofield[3], Dara Bobb-Semple[3], Jing Tao[4]
C. Jaye[5], D.Fischer[5], and Stanislaus S. Wong[3,4]

[1]Department of Physics, New Jersey Institute of Technology, Newark, NJ 07102
[2]Department of Physics and Astronomy, Rutgers University, Piscataway, NJ 08854
[3]Department of Chemistry, State University of New York at Stony Brook, Stony Brook, NY 11794
[4]Condensed Matter Physics and Materials Science Department, Brookhaven National Laboratory, Upton, NY 11973
[5]Materials Science and Engineering Laboratory, National Institute of Standards and Technology, Gaithersburg, MD 20899

To understand the nature of the valence state on Ti, O and Sr in the nanoparticles relative to bulk samples we conducted detailed spectroscopic and structural studies. In Fig. 2(c), we computed the derivative of the absorption spectra near the main line of the Ti K-Edge measurement (Fig. 2(a)) and showed that there is no significant chemical shift (shift of the inflection point) in the 10 nm sample compared with the bulk sample. Any shift is significantly below the 0.15 eV step used in data collection. We note that the Ti K-absorption edge inflection point positon is extremely sensitive to valence changes [1]. The Ti ionic state n relative to Ti metal is approximately given by $n = (0.28\pm0.01)*\Delta E$ (where $\Delta E$ is the chemical shift). Hence any valence change in Ti is below 0.05 units. We further examine the soft x-ray spectra at Ti L2 and L3 edges, the O K-Edge and the Sr L1-Edge (see Fig. S1). The Ti L2/L3-Edge spectra are composed of split Ti 3d $t_{2g}$, eg pairs. The second peak in the pair is sensitive to the Ti valence [2]. In this system, a shift of ~0.5 eV gives a valance change of 0.26 units. The upper limits on the valence shifts from these spectra are consistent with the Ti K-Edge limit. TheO K-Edge spectra (Fig. S1(b)) and Sr L1-Edge also show the same effect (no observable shifts) indicating that the $SrTiO_3$ nanoparticles are stoichiometric (no O, Sr or Ti defects) since the peak positions and relative intensities of the spectra do not change significantly. What is seen though is that amplitudes of spectra of the O K-Edge spectra decrease with reduction in particle size. We have also conducted total scattering x-ray diffraction measurements at NSLS X17A (Figs. S2 and S3) with high q-value to examine the atomic pair distribution functions and found no change in the Sr/Ti ratio between bulk and 10 nm samples. Comparison of the O K-Edge spectra of the 10 nm and 82 nm normalized at the Ti L2/L3-Edge (Fig. S4) also show no change in Ti:O ratio. Furthermore, we measured high-resolution synchrotron x-ray diffraction patterns (Fig. S5) and found that while the bulk sample is cubic with a = 3.9060(2) Å, the 10 nn sample exhibits a non-negligible tetragonal splitting with a = 3.9142(3) Å and c = 3.9118(6) Å. The results reveal an expansion of the lattice as the particle size is increased. (This expansion was seen previously [3] in low resolution laboratory diffraction measurements but here we have extracted a tetragonal splitting with the higher resolution data.) The lattice expansion with reduction in particle size is consistent with the reduction in the O K-Edge spectral intensity if we consider a reduction in hybridization of the O p and Ti d bands and O p with Sr d bands occurs as particle size is reduced (see Ref [4] for discussion of the electronic band in perovskites). We note that in the case of $BiFeO_3$, similar behavior is observed with particle size reduction [5]. However the mechanism behind the physical change in complex oxide nanoparticles with particle size needs to be addressed theoretically in future work. The weak distortions are more readily probed by local structural probes such as XAFS as seen in the text.

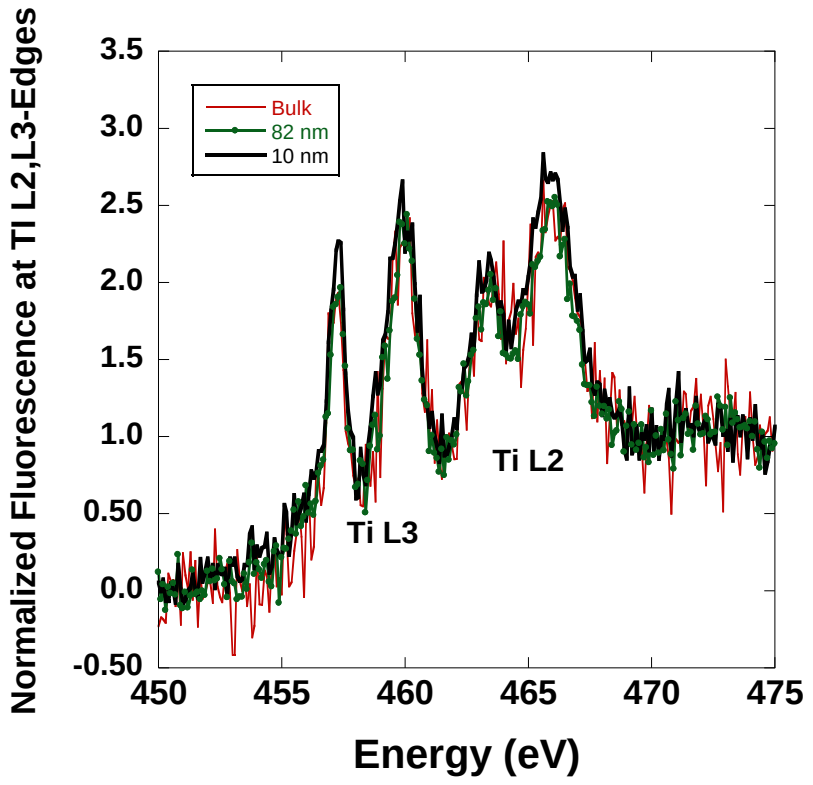

**(a)**

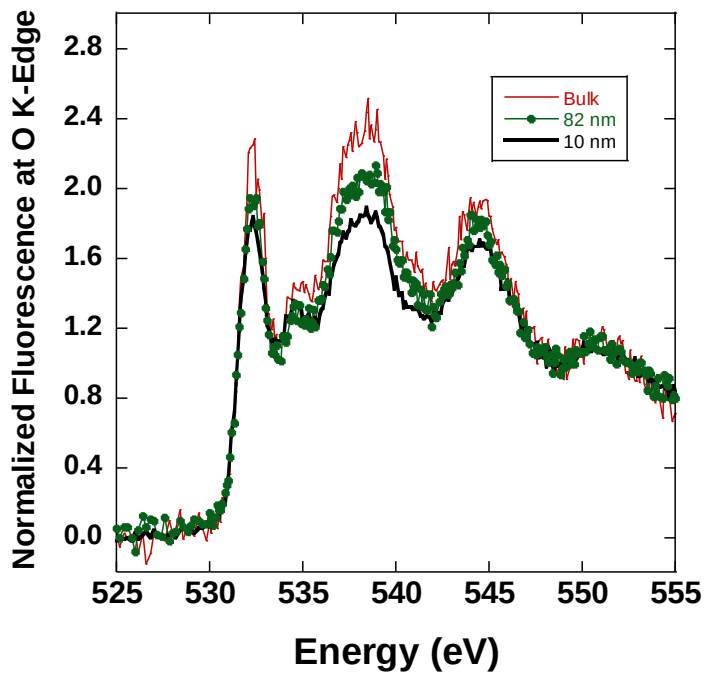

**(b)**

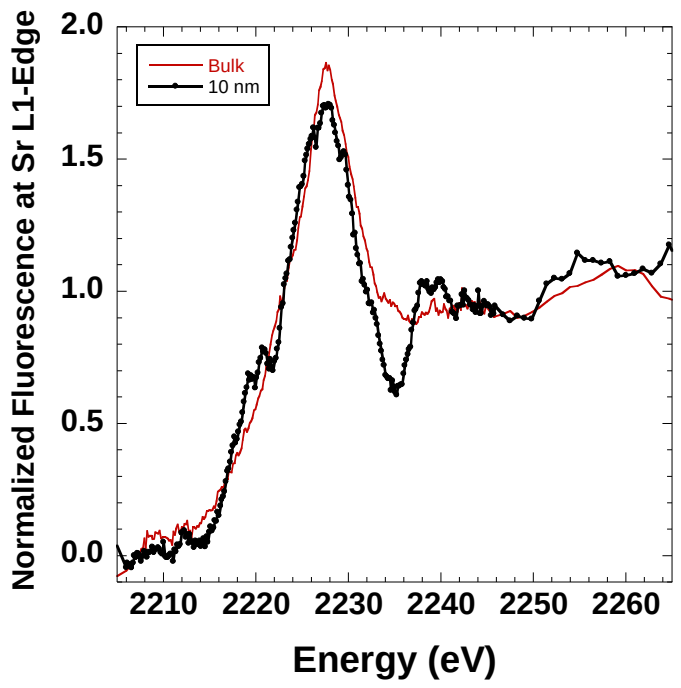

**(c)**

**Fig. S1.** Nanoparticle spectra at the (a) Ti L2,L3-Edge, (b) the O K-edge and (c) the Sr L1-Edge. No chemical shifts (valence changes) in the peak positions are observed relative to the bulk spectra for Ti, O and Sr. However, the O K-edge spectra show a reduction in amplitude with particle size suggesting reduction in hybridization of the O(p) with Ti(d) and with Sr(d) bands.

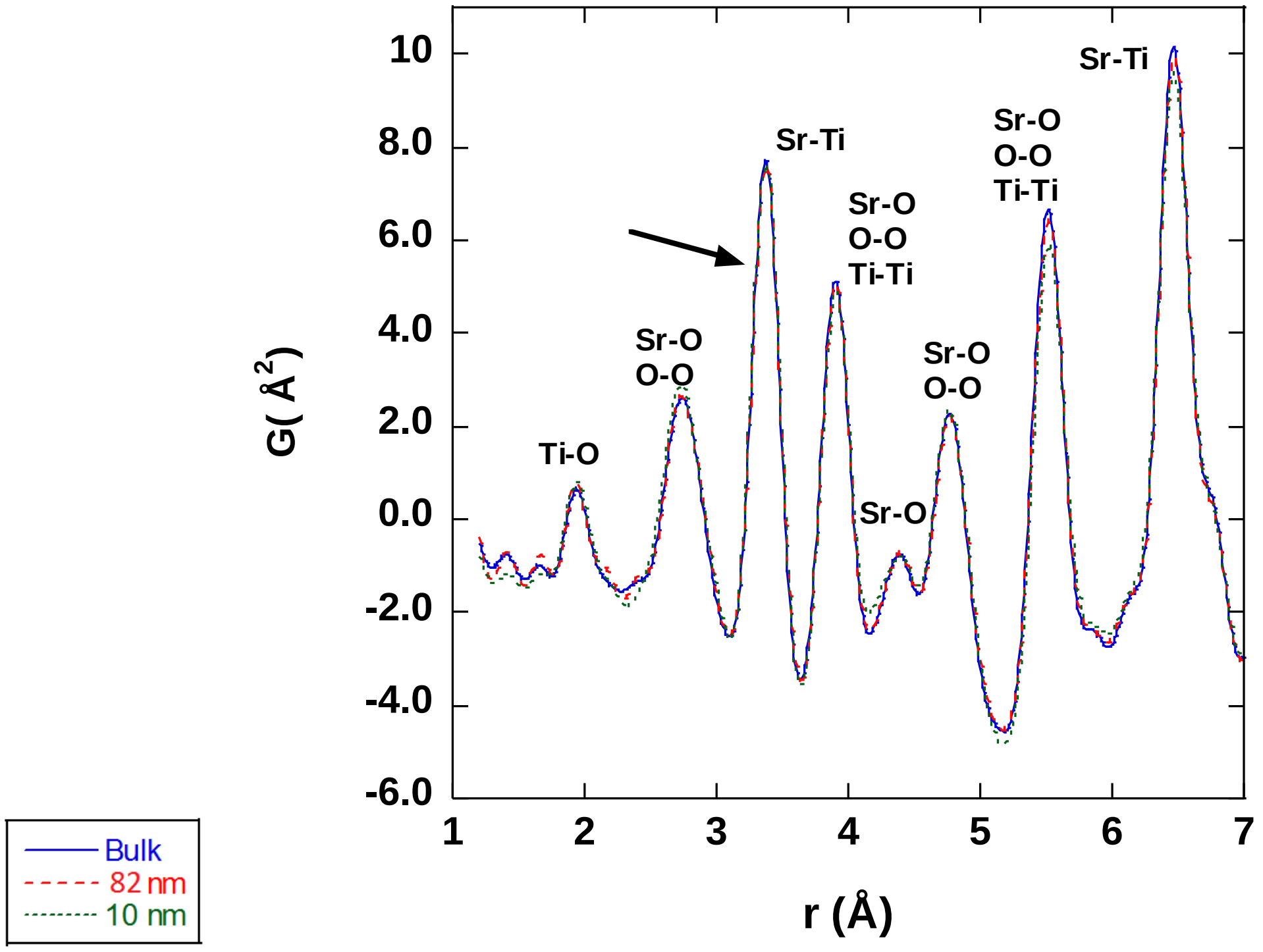

**Fig. S2.** Radial structure function of nanoparticles and bulk sample. The peak just above 3 Å is unmodified by the amplitude envelope due to particle size and indicates that the Sr:Ti ration is unity.

In Fig. S2. we show the x-ray derived structure function G(r). G(r) is the reduced atomic pair distribution function which oscillates about zero and is obtained directly from the scattering data, S(Q) [6]. The function $G(r)=\frac{2}{\pi}\int_0^\infty Q[S(Q)-1]\sin(Qr)dQ$ is related directly to the standard pair distribution function $g(r)$. The statistical variation in the data derive from examining multiple scans is at the level of the diference between the bulk and 82 nm sample (See Fig. S3). The Ti-Sr low-r peak which is not modified by the particle size envelope (see illustration of particle envelope function in Phys. Rev. B **76**, 115413 (2007)) is the same for all samples. This indicates that no significant change in the occupancy of the Sr or Ti sites occurs between the bulk and nanosacale samples. The peaks containing Sr-O and Ti-O pairs show variation possibly sensing the off-centering of the O atoms. However, neutron scatting measurements with high-q data would be better suited to extract this level of information for O sites.

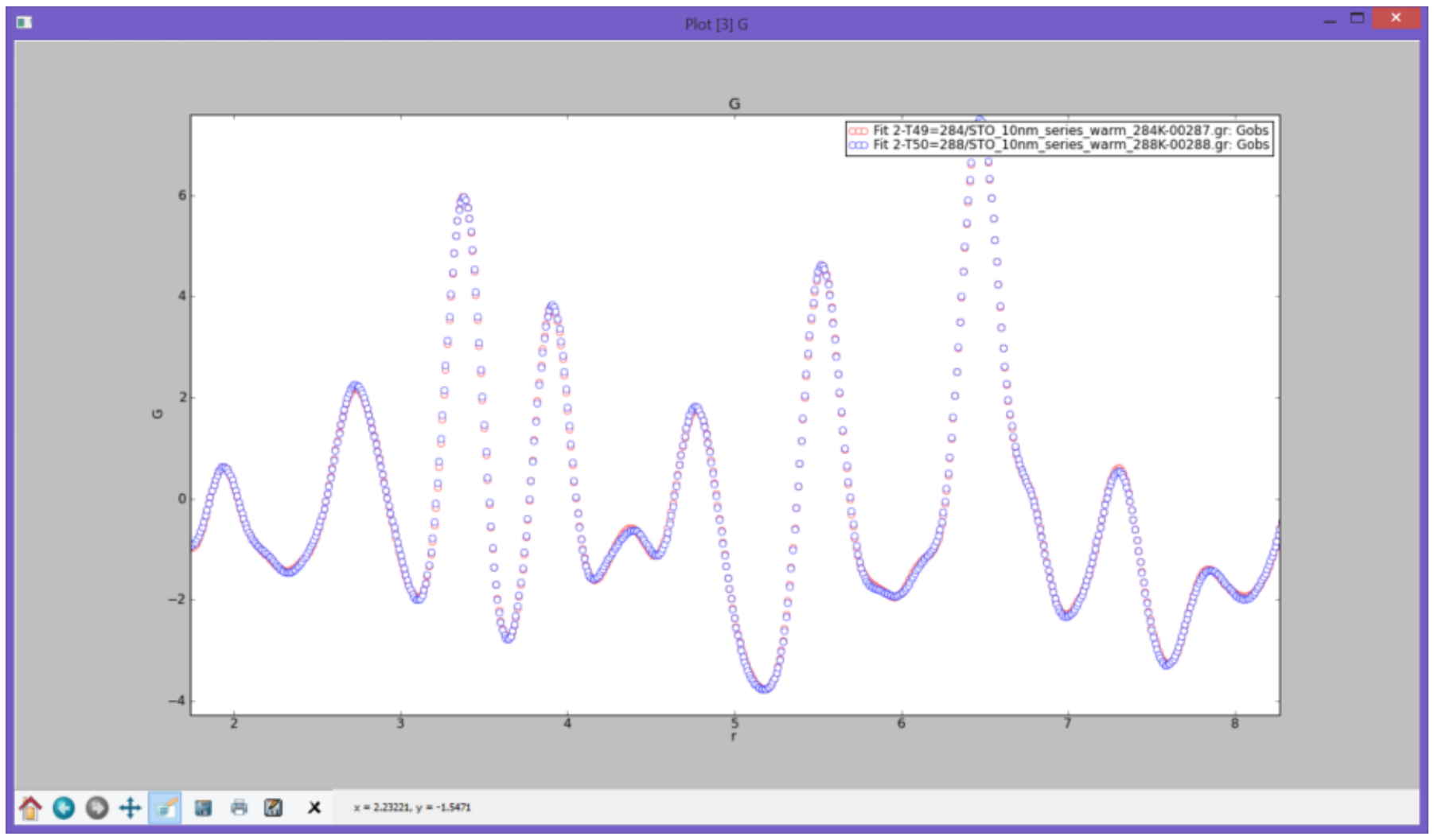

**Fig. S3** Two scans, one at 288 K and one at 284 K, for the 10nm sample showing the statistical spread in the data. This is typical of the data taken for the bulk and 82 nm samples also.

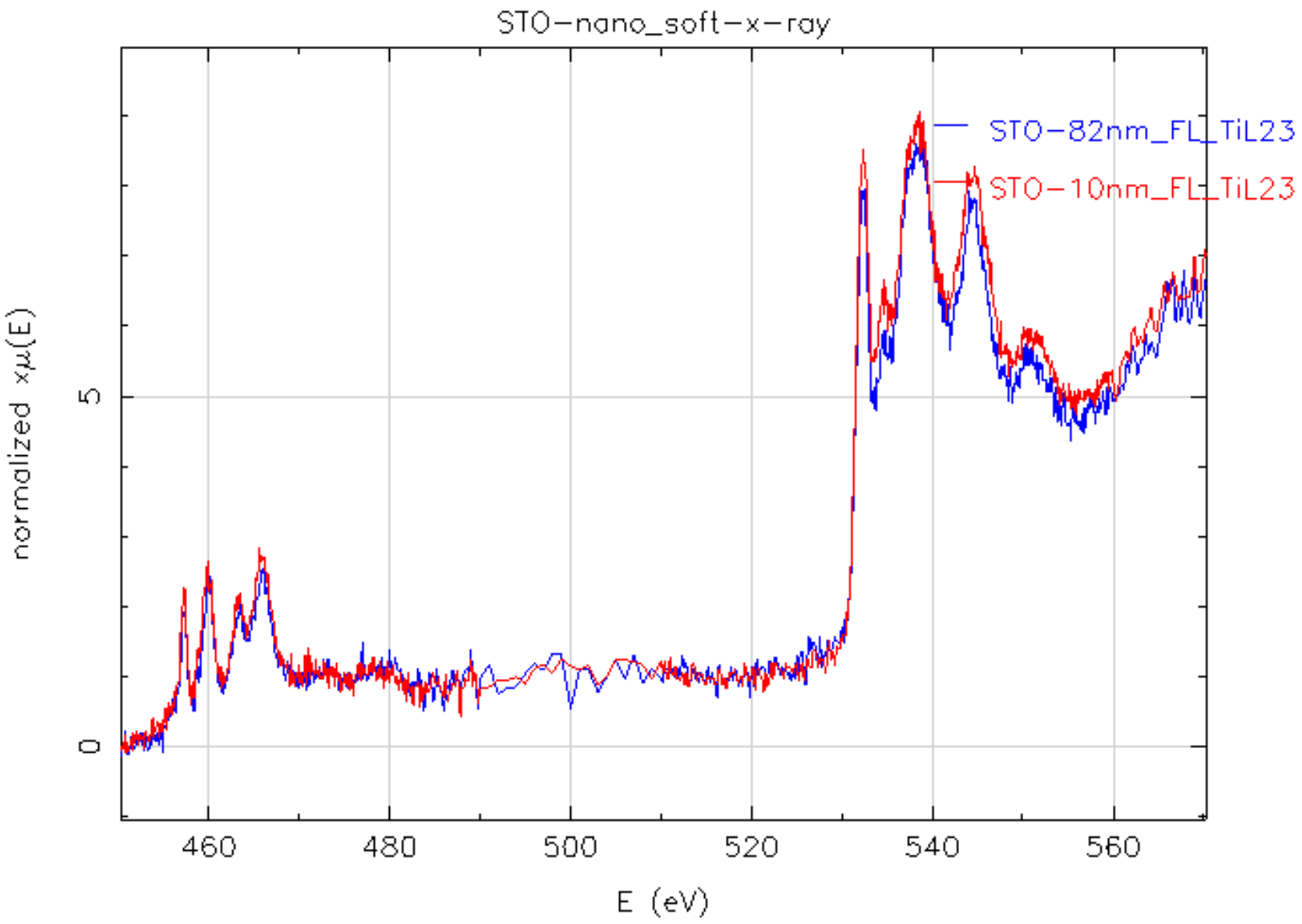

**Fig. S4.** The full spectra of 10 nm and 82 nm samples normalized at the Ti L2,L3 edge reveal a ratio of O to Ti that is the same for both samples (see for example Phys. Rev Lett **86**, 4056 (2001)).

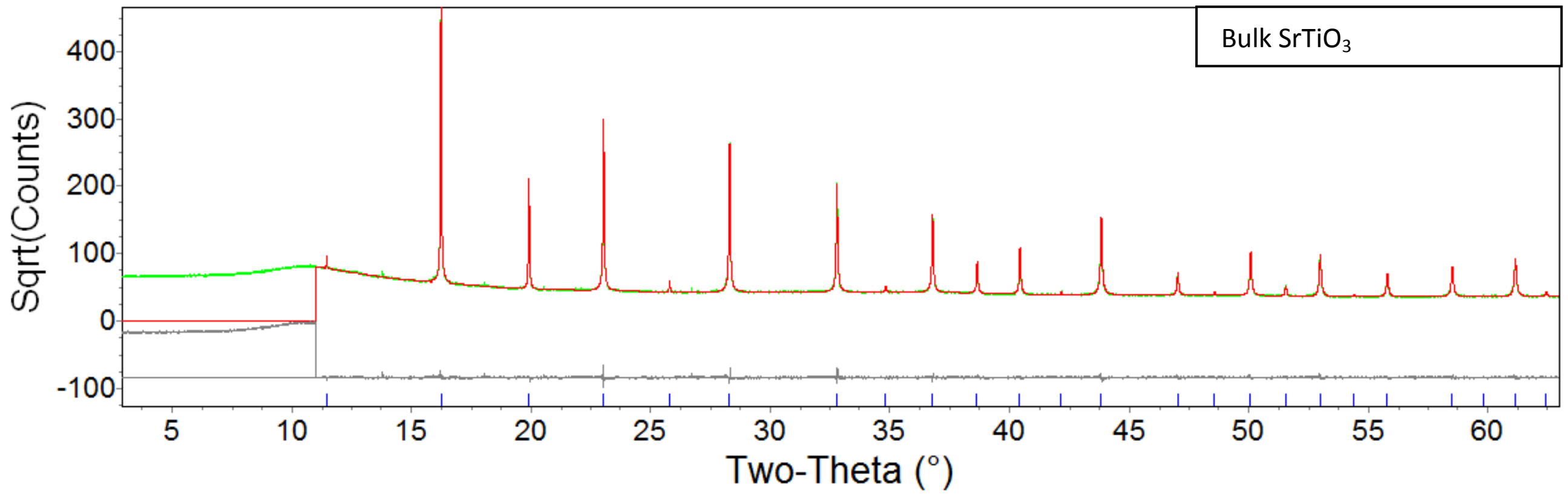

(a)

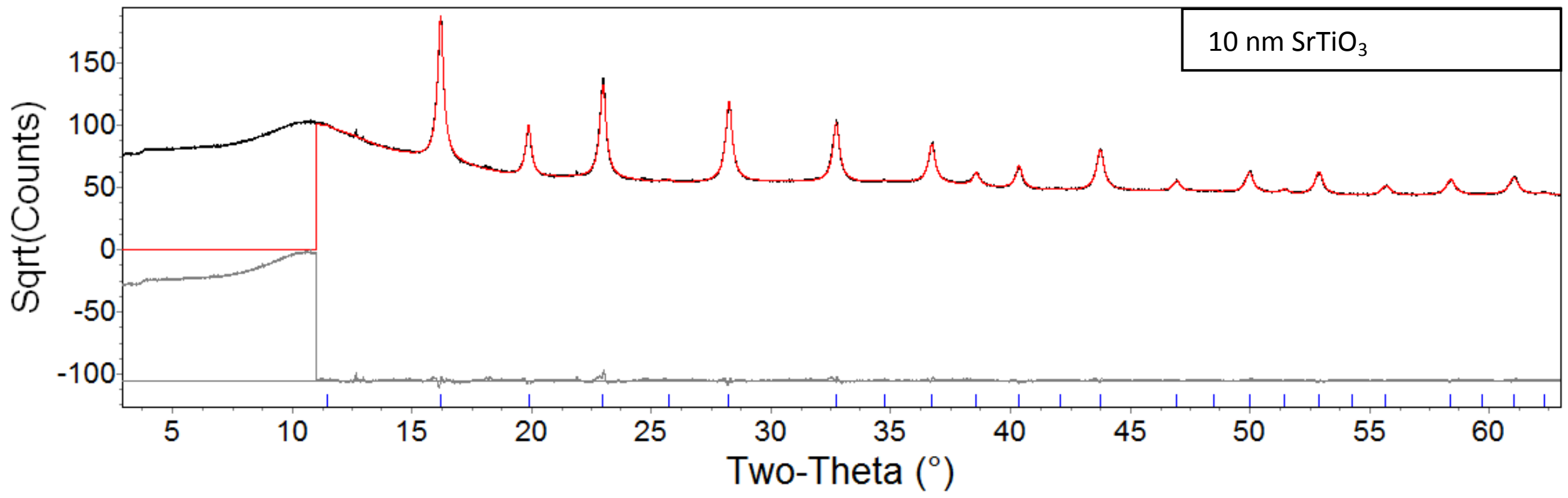

(b)

**Fig. S5.** Fit of XRD pattern for (a) bulk and (b) 10 nm samples excluding the region below 11 °. Weak unmatched peaks and the bump near 11 ° are from the quartz capillary containing the sample.

A NIST silicon standard was used to define the peak profiles used in the Rietveld refinements. The Topas refinement package was used for data analysis [7]. A weak non negligible tetragonal distortion with a difference between the a and c parameters of ~0.002 Å is found when fits are done for the nanoparticles to the P4mm space group. The weak distortions are more readily probed by local structural probes such as XAFS as seen in the main text.

## High Resolution X-Ray Diffraction Results

**Bulk ($R_{wp}$ = 3.90 %)**
a=3.9060 (2) Å

**10 nm ($R_{wp}$ = 2.39 %)**
a= 3.9142(3) Å
c= 3.9118(6) Å

## References

[1] J. Graetz, J. J. Reilly, J. Johnson, A. Y. Ignatov, and T. A. Tyson, Applied Physics Letters **85**, 500 (2004).

[2] D. A. Muller, N. Nakagawa, A. Ohtomoto, J. L. Grazul, and H. Y. Hwang, Nature **430**, 657 (2004).

[3] X. W. Wu, D. J. Wu, X. J. Liu, S. S. Comm. **145**, 255 (2005).

[4] (a) M. Abbate, F. M. F. de Groot, J. C. Fuggle, A. Fujimori, O. Strebel, F. Lopez, M. Domke, G. Kaindl, G. A. Sawatzky, M. Takano, Y. Takeda, H. Eisaki and S. Uchida, Phys. Rev. B **46**, 4511 (1992). (b) K. van Benthem, C. Elsasser, R. H. French, J. Appl. Phys. **90**, 6156 (2001)\

[5] F. Huang, Z. Wang, X. Lu, J. Zhang, K. Min, W. Lin, R. Ti, T. T. Xu, J. He, C. Yue, and J. Zhu, Sci. Rep. **3**, 2907 (2013), see Fig. 3.

[6] T. Egami and S. L. J. Billinge, Underneath the Bragg Peaks: Structural Analysis of Complex Materials, (Pergamon, Amsterdam, 2003).

[7] A. A. Coelho, J. Appl. Cryst. **33**, 899 (2000).